\documentclass[nofootinbib,preprintnumbers,amsmath,amssymb,onecolumn,unsortedaddress,superscriptaddress,twocolumn]{revtex4-1}
\usepackage[utf8]{inputenc}
\usepackage{graphicx, cancel, booktabs, tabularx, amsmath, amsfonts, amssymb, bm, hyperref, placeins}
\usepackage[normalem]{ulem}

\usepackage[table]{xcolor}

\renewcommand{\t}[1]{\text{#1}}
\renewcommand{\d}{\text{d}}

\renewcommand{\b}[1]{\mathbf{#1}}

\renewcommand{\vec}[1]{\mathbf{#1}}

\newcommand{\mean}[1]{\left\langle {#1} \right\rangle}

\newcommand{\halb}{\frac{1}{2}}

\newcommand{\sigmaA}{\sigma^\mathrm{A}}
\newcommand{\sigmaAOS}{\sigma^\mathrm{A}_\mathrm{OS}}
\newcommand{\sigmaAMD}{\sigma^\mathrm{A}_\mathrm{MD}}
\newcommand{\V}{\mathcal V (\vec R)}

\newcolumntype{L}[1]{>{\raggedright\arraybackslash}p{#1}}

\newcolumntype{C}[1]{>{\centering\arraybackslash}p{#1}}

\newcolumntype{R}[1]{>{\raggedleft\arraybackslash}p{#1}}

\begin{document}
\title{Anharmonicity Measure for Materials}
\author{Florian Knoop}
\author{Thomas A. R. Purcell}
\author{Matthias Scheffler}
\author{Christian Carbogno}
\affiliation{Fritz-Haber-Institut der Max-Planck-Gesellschaft, Faradayweg 4–6, D-14195 Berlin, Germany}

\begin{abstract}
Theoretical frameworks used to qualitatively and quantitatively describe nuclear dynamics in solids are often based on the harmonic approximation.
However, this approximation is known to become inaccurate or to break down completely in many modern functional materials.
Interestingly,  there is no reliable measure to quantify anharmonicity so far. Thus, a systematic classification of materials in terms of anharmonicity and a benchmark of methodologies that may be appropriate for different strengths of anharmonicity is currently impossible.
In this work, we derive and discuss a statistical measure that reliably classifies compounds across temperature regimes and material classes by their ``degree of anharmonicity''.
This enables us to distinguish ``harmonic'' materials, for which anharmonic effects constitute a small perturbation on top of the harmonic approximation, from strongly ``anharmonic'' materials, for which anharmonic effects become significant or even dominant and the treatment of anharmonicity in terms of perturbation theory is more than questionable.
We show that the analysis of this measure in real and reciprocal space is able to shed light on the underlying microscopic mechanisms,
even at conditions close to, e.g., phase transitions or defect formation.
Eventually, we demonstrate that the developed approach is computationally efficient and enables rapid \emph{high-throughput} searches by scanning over a set of several hundred binary solids.
The results show that strong anharmonic effects beyond the perturbative limit
are not only active in complex materials or close to phase transitions, but already at moderate temperatures in
simple binary compounds.
\end{abstract}
\date{\today}
\maketitle

\section{Introduction}
In condensed-matter physics and materials science,
the dynamics of nuclei plays a decisive role for many materials properties.
At the lowest level of theory, these vibrations are commonly described in the \emph{harmonic approximation},
in which the
potential-energy surface~$\mathcal V (\vec R)$ is approximated by $\mathcal{V}^{(2)} (\vec R)$,
a second-order Taylor expansion in the atomic displacements,
$\{\Delta \vec R_I\}$, about a minimum energy configuration,
$\{ \vec R_I^0 \}$, in terms of \emph{force constants},
$\Phi_{\alpha, \beta}^{I, J}$~\cite{BornHuang}.
The dynamical properties are determined by the model Hamiltonian
\begin{align}
	\mathcal{H}^{(2)} = \sum_I \frac{\vec P_I^2}{2 M_I}
    + \halb \sum\limits_{\genfrac{}{}{0pt}{2}{I , J}{\alpha, \beta} } \Phi_{\alpha, \beta}^{I, J} \,\Delta R_I^\alpha  \Delta R_J^\beta
	\label{eq:ha_1}
\end{align}
for the nuclear momenta, $\vec P_I$, and positions, $\vec R_I =  \vec R_I^0 + \Delta \vec R_I$.
Since this Hamiltonian
separates into $3N$ uncoupled harmonic oscillators, called \emph{phonons},
both the classical equations of motion and the quantum-mechanical Schrödinger equation can be solved analytically.
This
allows for the computation of thermodynamic properties such as
the harmonic free energy and the heat capacity~\cite{BornHuang, Dove}.
Since computing $\Phi_{\alpha, \beta}^{I, J}$ from first principles is a straightforward and computationally affordable task~\cite{Parlinski1997, phonopy, Plata2017}, the
harmonic approximation is a popular tool in materials science.

Material properties and phenomena that are described inaccurately or not at all within such a harmonic model are generally referred to as \emph{anharmonic} effects.
These effects
include i) the temperature dependence of equilibrium properties like thermal lattice expansion,
ii) thermal shift of vibrational frequencies and linewidth broadening, iii) phase transitions, and iv) heat transport. All of these properties either diverge or vanish in the harmonic approximation~\cite{Leibfried1961, Klemens1958, Dove, Fultz2010}.

In recent years, significant advances in modeling anharmonic effects were achieved,
especially in the field of thermal transport.
To date, anharmonic effects can be addressed either
exactly via non-perturbative \emph{ab initio} Molecular Dynamics~(aiMD)~\cite{Carbogno2016, Marcolongo2016},
or, more commonly, with approximate, perturbative models~\cite{Broido2005, Esfarjani2011, Tadano2014, Hellman2014, Tadano2015, Feng2017, Ravichandran2018, Simoncelli2019}.
In the latter case, the Taylor expansion of the potential~$\mathcal V ( \vec R)$ is extended
beyond the second order and the additional terms are treated as a perturbation of the harmonic model.
Yet, there is a fundamental question about the applicability of perturbative approaches:
The reliability of any perturbation expansion is controlled by the strength of the perturbation, which has to be ``small'' with respect to the reference~\cite{Altland2010}.
However, no measure exists to date to reliably
quantify the strength of anharmonic effects. First, this lack
prevents the systematical and quantitative exploration of the applicability limits of perturbative techniques.
Second, this hinders a systematic classification of materials across temperature regimes by anharmonicity.
Given that anharmonic effects may influence
or even determine macroscopic material properties, understanding qualitative trends that govern anharmonicity across material space is a challenge of growing importance.

In this work, we address this open issue by deriving and validating the required \emph{anharmonicity measure}.
As discussed below, it
(a)~allows for a systematic and quantitative
classification of compounds across material space
from systems with only mild anharmonic contributions, to strongly anharmonic systems where the phonon picture is invalid,
(b)~establishes a link between the actuating microscopic mechanisms and macroscopic properties,
and (c)~requires a fraction of the computational cost of either perturbative  or aiMD calculations
and thus paves the way for \emph{high-throughput} anharmonicity classification of materials.
After presenting and discussing the underlying theory, definitions, and the numerical aspects for two exemplary materials in Sec.\,\ref{sec:Anharmonicity}-\ref{sec:Sampling},
we show how our anharmonicity measure
can be incorporated into an \emph{ab initio}
high-throughput screening of material space in Sec.\,\ref{sec:Screening}.
We discuss our findings and their implications for theoretical solid-state physics and materials science in Sec.\,\ref{sec:conclusion}.

\section{Definition of Anharmonicity}
\label{sec:Anharmonicity}
In the Born-Oppenheimer approximation~\cite{Born1927},
the dynamics of a nuclear many-body system is described by the Hamiltonian determined by the potential energy $\mathcal{V}(\vec{R})$,
\begin{align}
	\mathcal{H}(\vec R, \vec P)
	= \sum_I \frac{\vec P_I^2}{2 M_I}  + \mathcal{V}(\vec{R})~,
	\label{eq:H1}
\end{align}
where \mbox{\(\vec{R} = \{\vec{R}_1, \ldots, \vec{R}_N\}\)}
denotes the atomic coordinates and
\mbox{$\vec{P} = \{\vec{P}_1, \ldots, \vec{P}_N\}$}
the respective momenta.
Treating the nuclei as classical particles, the Hamiltonian generates
equations of motion for each atom $I$,
\begin{align}
M_I \, \ddot{\vec R}_I (t)
= \vec F_I (t)
= - \b{\nabla}_I \mathcal{V} \left(\vec R (t) \right)~,
\label{eq:Newton}
\end{align}
with atomic mass $M_I$, acceleration $\ddot{\vec R}_I$, and the force $\vec F_I$. We note in passing that the formalism and calculations presented in this work focus on the classical limit,~i.\,e.,~we neglect low-temperature nuclear quantum effects. This is justified in the investigated temperature range above 300\,K for the inorganic solids discussed here. A generalization to a full quantum treatment of the nuclei~\cite{Markland2018} is possible and could be used to investigate anharmonic effects in organic materials~\cite{Litman2019}, but would go beyond the scope of this work.

As mentioned in the introduction, the \emph{harmonic approximation} replaces the many-body potential $\V$ by a second-order Taylor expansion around equilibrium~$\vec R^0$,
\begin{align}
\mathcal{V}^{(2)}(\vec{R} = \vec{R}^0 + \Delta \vec R)
= \frac{1}{2} \sum_{\substack{I, J \\ \alpha, \beta}}
\Phi_{\alpha, \beta}^{I, J} \, \Delta R_I^\alpha \Delta R_J^\beta ~,
\label{eq:TEP1}
\end{align}
in terms of the force constants
\begin{align}
\Phi_{\alpha, \beta}^{I, J}
= \left. \frac{\partial^2 \mathcal V ({\bf R})}{\partial R_I^\alpha \partial R_J^\beta} \right|_{\vec R^0} ~.
\label{eq:TEP2}
\end{align}
Accordingly, the forces are approximated as
\begin{align}
	F^{(2)}_{I, \alpha} = - \sum_{J, \beta} \Phi_{\alpha, \beta}^{I, J} \, \Delta R_{J}^\beta~.
	\label{eq:F2}
\end{align}
On this approximated potential-energy surface, the equations of motion can be solved analytically~\cite{Dove}.
Differences between the actual and the harmonic nuclear dynamics arise from the difference between the exact potential~$\mathcal{V}(\vec{R})$ and the harmonic potential~$\mathcal V^{(2)}(\vec{R})$, i.\,e.,~from anharmonic contributions.
Consequently, we define the \emph{anharmonic contribution to the potential} as
\begin{align}
	\mathcal{V}^\t{A} (\vec R)
	\equiv \mathcal{V}(\vec R) - \mathcal{V}^{(2)} (\vec R) ~.
	\label{eq:Va}
\end{align}

\begin{figure}
	\centering
	\includegraphics{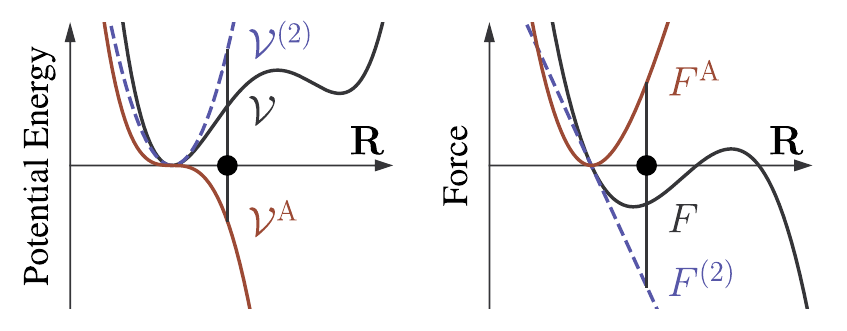}
	\caption{Left: Sketch of a one-dimensional potential-energy surface $\V$ (solid black), its harmonic approximation $\mathcal{V}^{(2)} (\b R)$ (dashed blue), and the anharmonic contribution $\mathcal{V}^{\rm A} (\b R)$ (solid red). Right: The force ${F} ({\bf R})$ given by the derivative of the potential energy $\V$ (black), the force $F^{(2)}$ stemming from the harmonic potential $\mathcal{V}^{(2)} (\b R)$ (blue), and the anharmonic contribution $F^\mathrm{A} = F - F^{(2)}$ (red), cf.~Eq.\,\eqref{eq:FA}.}
	\label{fig:PES}
\end{figure}

Analogously, the \emph{anharmonic contributions to the forces} that enter the equations of motion are given by
\begin{eqnarray}
\label{eq:FA}
F_{I, \alpha}^{\t A} (\vec R)
= F_{I, \alpha} (\vec R) - F_{I, \alpha}^{(2)} (\vec R)~,
\end{eqnarray}
with $\b F_I^{(2)}$ obtained from Eq.\,\eqref{eq:F2}. The construction is qualitatively depicted in Fig.\,\ref{fig:PES}.
Compared to the potential $\V$, working with the forces has the advantage that they naturally decompose into $3N$ components,
hence giving
straightforward access to a real-space per-atom and a reciprocal-space per-mode analysis.
For the latter purpose, we construct the dynamical matrix
\begin{eqnarray}
D_{\alpha, \beta}^{\tilde I, \tilde J} (\vec q)
~=~ \frac{1}{\sqrt{M_I M_J}}
\sum_L \t e^{\t i \vec q \cdot \vec R_L} ~ \Phi_{\alpha, \beta}^{\tilde I, J(\tilde J, L)}~.
\label{eq:Dynmat_q}
\end{eqnarray}
Here, $\tilde I$ labels the atoms in the primitive unit cell, $L$ denotes the Bravais lattice points,
$J(\tilde J, L)$ labels the periodic images of~$\tilde J$,
and $\vec q$ is the phonon wave vector~\cite{BornHuang}.
Diagonalizing~$D(\vec q)$ yields the harmonic \emph{eigenfrequencies}~$\omega_n(\vec q)$ and \emph{eigenmodes}~$\vec e_{n}(\vec q)$.
For readability, we use the generalized index $s = (\vec q, n)$ in the following.
In this notation,
the mode-resolved forces are
\begin{eqnarray}
	F_{s} = \sum_I \frac{1}{\sqrt{M_I}} ~ \vec e_{sI} \cdot \vec F_I~.
	\label{eq:Fs}
\end{eqnarray}

\section{Quantifying Anharmonicity}
\label{sec:AnharmonicityMeasure}
\begin{figure}
	\centering
	\includegraphics{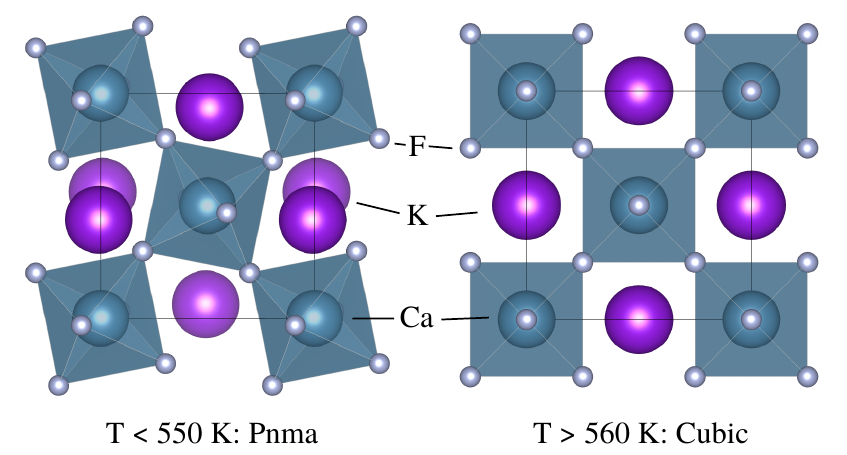}
	\caption{\label{fig:KCaF3}
		KCaF$_3$ in the low-temperature Pnma~(left) and high-temperature aristotype phase~(right). Both structures are viewed along the long $b$-axis.~\cite{Momma2008}}
\end{figure}
\begin{figure}[h!]
	\centering
	\includegraphics{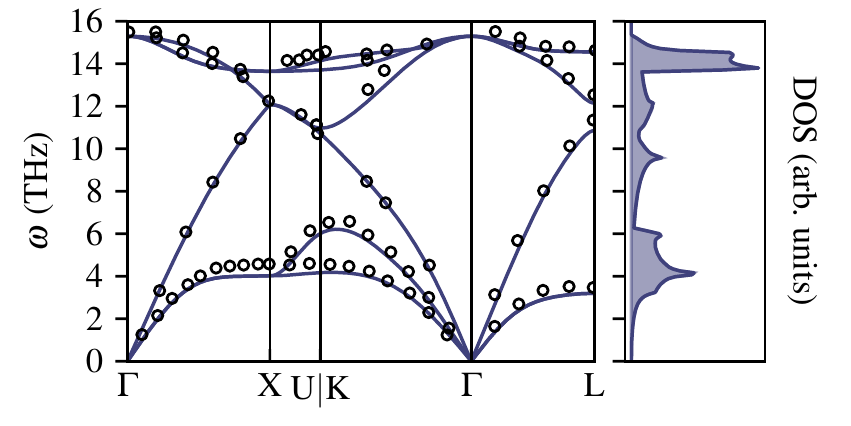}
	\caption{Phonon bandstructure of fcc-diamond silicon obtained for a supercell with 216 atoms. Open circles denote experimental reference data from inelastic neutron scattering at room temperature~\cite{Nilsson1972}.}
	\label{fig:bandstructure_Si}
\end{figure}
\begin{figure}[h!]
	\centering
	\includegraphics{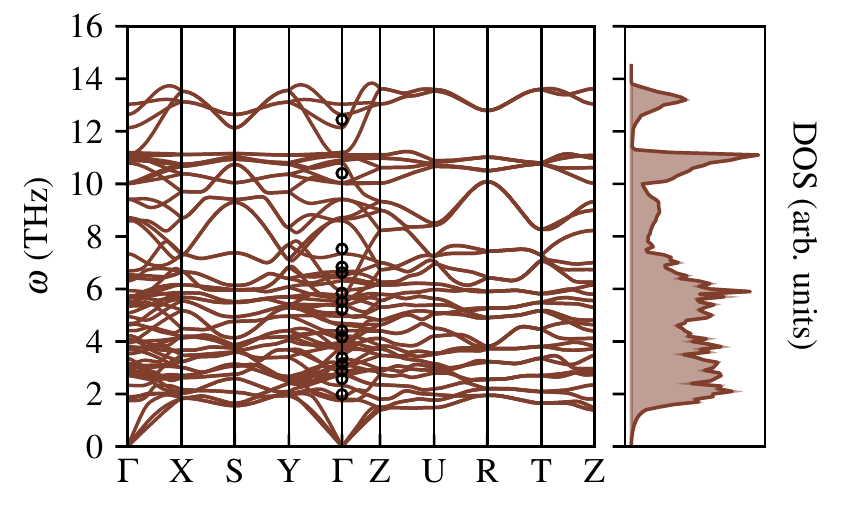}
	\caption{Phonon bandstructure of KCaF$_3$ in the Pnma structure obtained from a supercell with 160 atoms. Open circles denote experimental reference data from Raman scattering at 40\,K~\cite{Daniel1997}.}
	\label{fig:bandstructure_KCaF3}
\end{figure}

Two prototypical materials are used as examples in the following to elucidate the concepts developed in this work:
Silicon~(fcc-diamond) serves as example for a largely harmonic material, whereas the low-temperature structure of the perovskite KCaF$_3$~({Pnma}~\cite{Knight2005}, cf.~Fig.\,\ref{fig:KCaF3}),
is used as example for
a complex, strongly anharmonic material.
As shown in Fig.\,\ref{fig:bandstructure_Si}~and~\ref{fig:bandstructure_KCaF3}, both materials exhibit vibrational spectra of roughly the same frequency range in the harmonic approximation.
The perovskite KCaF$_3$ features an anion octahedral tilting~($a^- b^+ a^-$ in the Glazer notation~\cite{Glazer1972, Bulou1980}) which reduces with temperature, finally leading to a dynamically stabilized cubic crystal at 560\,K~\cite{Bulou1980, Demetriou2005}.
We will discuss this phase transition and the implications for anharmonicity quantification in more detail in Sec.~\ref{sec:DynamicallyStabilized}.

We investigate both compounds at room temperature via
{\it ab initio} molecular dynamics (aiMD) simulations~\cite{Car1985}
 at the GGA level of theory using the PBEsol exchange-correlation functional~\cite{PBEsol}, light~default basis sets~\cite{FHI-aims}, and a Langevin thermostat~\cite{Tuckerman} to perform canonical sampling in supercells of 216 atoms (Si) and 160 atoms (KCaF$_3$), respectively.
The aiMD is performed with a time step of 5\,fs for an initial thermalization period of 2\,ps and a sampling period of 8\,ps.
The chosen numerical settings ensure that all quantities of interest,~i.e., structural parameters such as lattice constants, dynamical properties such as vibrational frequencies, as well as thermodynamic averages such as the $\sigmaA$~measure introduced below are converged within $\pm 1$\,\%.

All the necessary calculations and tools used for performing the calculations and investigating the results
are implemented in our python
package \textit{FHI-vibes}~\cite{vibes}. It builds on top of the
\emph{Atomistic Simulation Environment}
(\textsc{ASE})~\cite{HjorthLarsen2017}, interfaces with \textit{phonopy} for building harmonic force constants~\cite{phonopy}, and integrates tightly with the all-electron, numeric atomic orbitals code \textit{FHI-aims} for performing \emph{ab initio} calculations of energy, forces, and stress~\cite{FHI-aims, Knuth2015}.

\subsection{Normalization of Forces}
\label{sec:Fnorm}
A prerequisite for comparing forces acting in different systems under various thermodynamic conditions is that these forces are normalized. For this purpose,
we characterize each force component $F_{I, \alpha} (t)$ observed during the simulation by
the probability-density function~$p_{{I, \alpha}} (F)$, and use the definition
of the thermodynamic average to obtain
\begin{align}
\mean{F_{I, \alpha}} &= \int_{-\infty}^\infty F \, p_{{I, \alpha}} (F) \,\d F ~,
	\label{eq:meanF}\\
	p_{{I, \alpha}} (F) &= \frac{1}{N_t} \sum_{t} \delta \left( F - F_{I, \alpha} (t) \right) ~,
	\label{eq:pIa}
\end{align}
where $\delta (F)$ denotes the delta distribution.
To characterize the whole system, we use the \emph{mixture probability distribution}
\begin{align}
	p (F)
	= \frac{1}{3N_I} \sum_{I, \alpha} p_{I, \alpha} (F)~,
	\label{eq:pF}
\end{align}
i.\,e.,~the weighted sum of probability distributions for each force component $F_{I, \alpha}$.
\begin{figure}[h]
	\centering
	\includegraphics{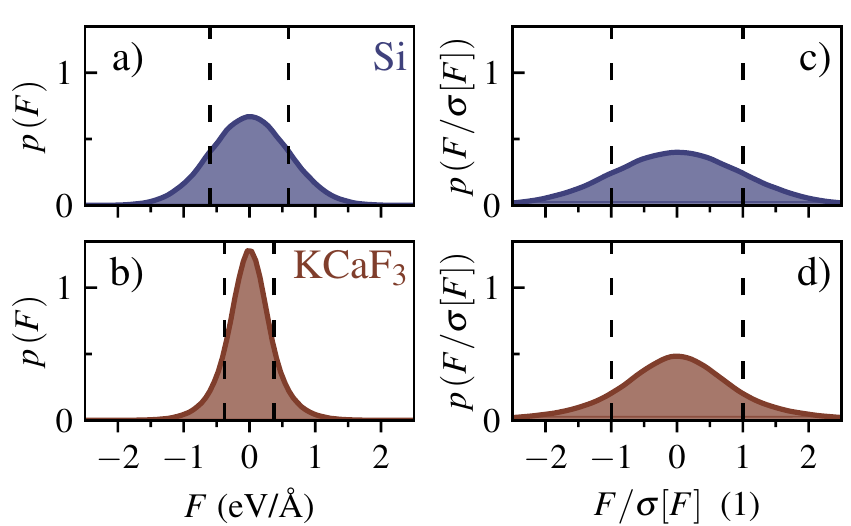}
	\caption{\label{fig:histogram_forces}
		Distribution of forces~$p (F)$ observed during aiMD simulations at 300\,K
		before~(left) and after normalization~(right) for silicon~(upper)
		and KCaF$_3$~(lower row). Dashed lines denote the standard deviation of the distribution.
	}
\end{figure}
Since the average of the individual force components vanishes in the absence of an external force,
$\mean{F_{I, \alpha}} = 0$,
the distribution $p(F)$ has a mean of zero. However, the width of~$p(F)$ depends on the actual material, as shown in plots of $p (F)$ for silicon and KCaF$_3$ in Fig.\,\ref{fig:histogram_forces}~a)~and~b).
We evaluate the width of the force distribution by computing its standard deviation,
\begin{align}
	\sigma [F] = \sqrt{\int_{-\infty}^\infty F^2 \, p(F) ~\d F}
	= \sqrt{\frac{1}{3 N_I} \sum_{I, \alpha} \left\langle F^2_{I, \alpha} \right\rangle}~,
	\label{eq:sigmaF}
\end{align}
with the thermodynamic average
\begin{align}
	\langle F_{I, \alpha}^2 \rangle
	= \frac{1}{N_t} \sum\limits_{t=1}^{N_t} F^2_{I, \alpha} (t)~.
	\label{eq:meanF2}
\end{align}
The analysis reveals that the distribution of forces in silicon exhibts a width of $\sigma [F_\text{Si}] = 0.60$\,eV/\AA, while KCaF$_3$ features a width of $\sigma [F_\text{KCaF}] = 0.38$\,eV/\AA. This is consistent with the phonon dispersions in Fig.\,\ref{fig:bandstructure_Si}~and~\ref{fig:bandstructure_KCaF3}, since KCaF$_3$ features more low-energy states, resulting, on average, in smaller restoring forces compared to silicon.
We take $\sigma\left[ F \right]$ as a measure for the average magnitude of forces acting in a material in thermodynamic equilibrium, including harmonic and anharmonic contributions. The average force $\sigma [F]$ therefore defines a scale in which the forces can be given and compared independent of the system by defining the \emph{normalized force}
\begin{equation}
 \vec{F}_I(t) \longrightarrow \vec{{F}}_I(t) \,/\, \sigma\left[ F \right] ~.
 \label{eq:Fnormalization}
\end{equation}
As shown in Fig.\,\ref{fig:histogram_forces}~c)~and~d), the mixture distributions of these normalized forces,
$p \left(F / \sigma [F] \right)$,
exhibit the same unit width
for both Si and KCaF$_3$.  This normalization thus allows to perform a meaningful comparison between the two materials.

\subsection{Anharmonicity Measure}
\label{sec:sigma}
To compute the anharmonic contribution to the forces, we use the aiMD forces~$\vec{F}_{I}(t)$ and obtain their harmonic contribution~$\vec{F}_{I}^{(2)}(t)$ by evaluating Eq.\,\eqref{eq:F2}
using the displacements $\Delta \vec R (t) = \vec R (t) - \vec R^0$ observed along the MD trajectory. The anharmonic force is then given by~$\vec{F}_{I}^\t{A}(t) = \vec{F}_{I}(t) - \vec{F}_{I}^{(2)}(t)$ as defined in~Eq.\,\eqref{eq:FA}.
In close analogy to the previous section, we use the probability distribution~$p_{I, \alpha} ({F}^\mathrm{A})$ and the mixture probability distribution~$p ({F}^\mathrm{A})$ to characterize the statistical behavior of~${F}^\mathrm{A}$.
Likewise,
we normalize~$\vec{F}^\t{A}_{I,\alpha}$ with respect to the force scale $\sigma [F]$.
Accordingly,
$\vec{{F}}_I^\t{A}(t) \,/\, \sigma\left[ F \right]$
describes the anharmonic contribution to the force on atom $I$ with respect to the average magnitude of the  total forces $\b F_I$.

\begin{figure}
 	\centering
 	\includegraphics{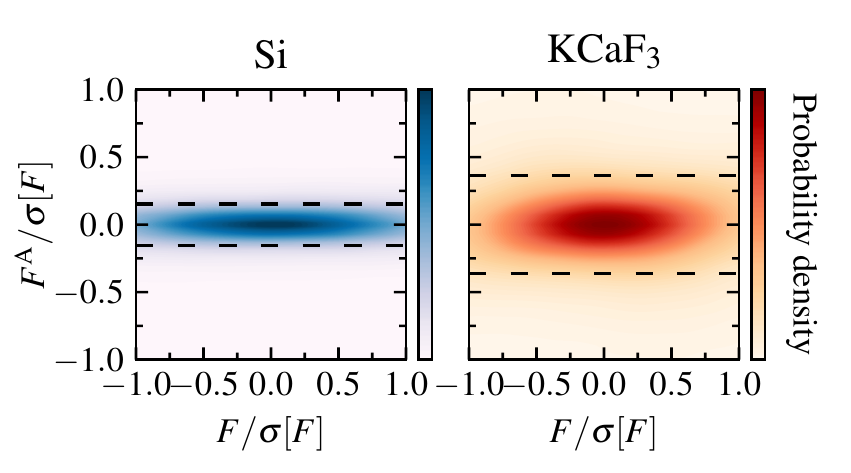}
  	\caption{\label{fig:histogram}
  		Joint probability densities $p(F, F^\mathrm{A})$ to find a force component $F$ and its anharmonic contribution $F^\mathrm{A}$ in units of
  		$\sigma [F]$.
  		Dashed lines denote the respective standard deviations. The color intensity increases linearly from zero to the maximum value of the respective distribution.}
 \end{figure}

For both silicon and KCaF$_3$ at 300\,K, the distributions~$p ({F}^\mathrm{A}/\sigma\left[ F \right])$ are plotted together with~$p ({F}/\sigma\left[ F \right])$ in Fig.\,\ref{fig:histogram} as joint probability plots.
Given that the force scale~$\sigma\left[ F \right]$ introduced above is used, the normalized forces are similarly distributed in x-direction
for both materials as discussed earlier.
On the y-axis, however, the distribution of the rescaled anharmonic forces
${F}^\mathrm{A} / \sigma[F]$
is significantly different for the two materials. In silicon,
the distribution is sharply peaked around~0 with a width of 0.15~$\sigma\left[ F_\text{Si} \right]$. From the distribution, we can quantify that only 15\,\% of the forces stem from anharmonic contributions on average and the probability of finding anharmonic force contributions of 0.5~$\sigma\left[ F_\text{Si} \right]$ or larger~is~$<0.01\,\%$.
This confirms the general understanding that silicon is largely harmonic and strong anharmonic contributions are essentially absent.
Conversely, the distribution of anharmonic forces for the perovskite KCaF$_3$ in Fig.~\ref{fig:histogram}~(right) is much broader, featuring a width of 0.36~$\sigma\left[ F_\text{KCaF} \right]$,~i.\,e., 36\,\% of the forces acting on the nuclei stem from anharmonic effects on average.
More importantly, finding strongly anharmonic force contributions of 0.5~$\sigma\left[ F_\text{KCaF} \right]$ or larger is $\simeq 16.5\,\%$, thus several orders of magnitude more probable than in silicon.
This means that these contributions are indeed significant in this compound, as argued above in the introduction of Sec.\,\ref{sec:AnharmonicityMeasure}.

In spirit of this discussion, we define the following measure for the quantitative estimation of the \emph{degree of anharmonicity} in a material:
\begin{eqnarray}
\sigmaA (T)
\equiv \frac{\sigma \left[{F}^\t{A} \right]_{T}}{\sigma \left[ {F } \right]_{T}}
=\sqrt{\frac{ \sum\limits_{I, \alpha} \left\langle \left( F_{I, \alpha}^\mathrm{A} \right)^2 \right\rangle_{T}}{
		\sum\limits_{I, \alpha} \left\langle \left( F_{I, \alpha} \right)^2 \right\rangle_{T}}} ~,
\label{eq:sigmaA}
\end{eqnarray}
with the thermodynamic ensemble average $\langle \cdot \rangle_{T}$ obtained according to Eq.\,\ref{eq:meanF2}.
$\sigmaA (T)$ measures the \emph{standard deviation of the distribution of anharmonic force components}
${{F}}^\t{A}_{I, \alpha}$ obtained from the \emph{ab initio} forces $\vec F$ and their harmonic approximation $\vec F^{(2)}$ according to Eq.\,\eqref{eq:FA}, normalized by the standard deviation of the \emph{ab initio} force distribution in the absence of external forces. This is mathematically equivalent to the \emph{root mean square error} (RMSE) of the harmonic model divided by the standard deviation of the force distribution.

\subsection{Atom- and Mode-Resolved Anharmonicity}
\label{sec:sigmas}
\begin{figure}
 	\centering
 	\includegraphics{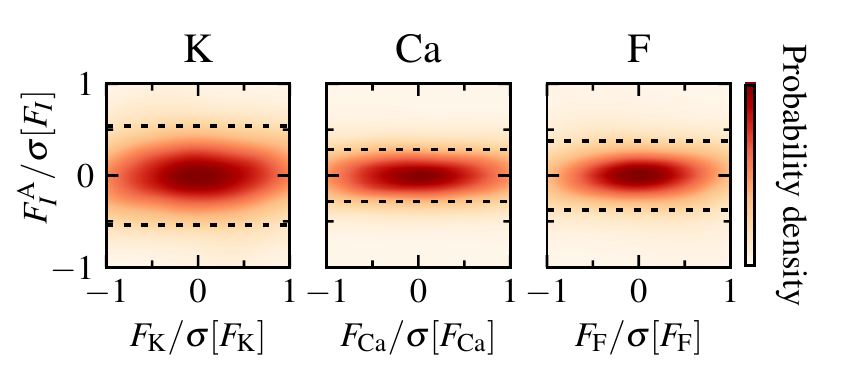}
 	\caption{\label{fig:histogram_KCaF3}
 		Joint probability densities to find a force component $F$ and its anharmonic contribution $F^\mathrm{A}$ in units of
 		$\sigma [F]$ for each atom type in KCaF$_3$ at 300\,K.
		$\sigmaA_\text{K} = 0.54$, $\sigmaA_\text{Ca} = 0.28$, $\sigmaA_\text{F} = 0.38$. The color saturation increases linearly from zero to the maximum value.
 	}
 \end{figure}

To estimate the degree of anharmonicity in a specific subset of the available degrees of freedom, denoted by $X$, we evaluate
	\begin{eqnarray}
	\sigmaA_X (T)
=\sqrt{\frac{ \sum\limits_{x \in X}\left\langle \left( F_{x} - F_{x}^{(2)} \right)^2 \right\rangle_{T}}{
			\sum\limits_{x \in X} \left\langle \left(F_{x} \right)^2 \right\rangle_{T}}} ~,
	\label{eq:sigmaAx}
	\end{eqnarray}
where $X$ can be,~e.\,g., a specific atom $I$, a group of atoms, or a vibrational mode $s$.
It is important to note that in Eq.\,\eqref{eq:sigmaAx}, we normalize by the width of the force components of interest, $\sigma \left[ {F_X} \right]$, and \emph{not} by the force scale $\sigma \left[ F \right] = \sqrt{\sum_X \sigma [F_X]^2}$ that averages over all available components, as in Eq.\,\eqref{eq:sigmaA}. By this means, we assess the \emph{relative} importance of anharmonicity in the degree(s) of freedom of interest.
As an example, Fig.\,\ref{fig:histogram_KCaF3} shows the species-resolved anharmonicities for KCaF$_3$. This analysis reveals that the forces acting on the K atoms have the largest anharmonic contribution, while the Ca and F atoms are more harmonic. This is a result of the K atoms moving in a relatively shallow potential
that allows them to participate significantly in the octahedral tilt, whereas the calcium atoms occupy the relatively stable vertices of the unit cell and its center,~cf.~Fig.\,\ref{fig:KCaF3}.

\begin{figure}[h!]
 	\centering
 	\includegraphics{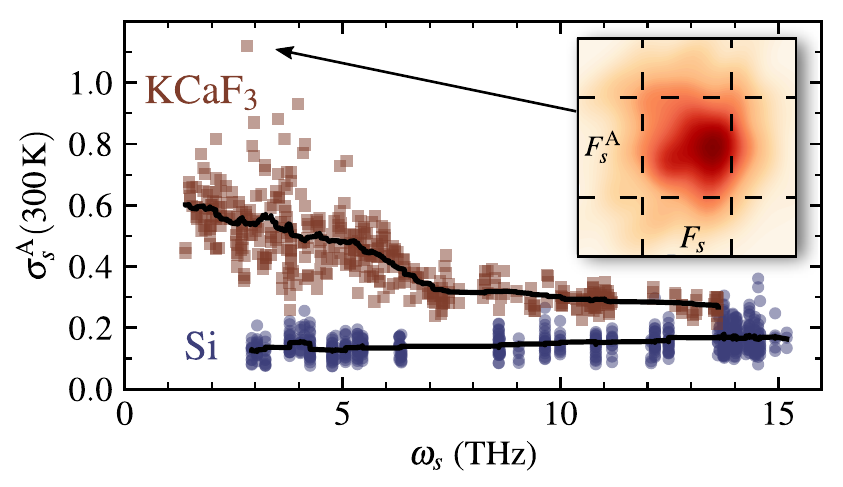}
  	\caption{\label{fig:sigma_mode_freq} Mode resolved degree of anharmonicity $\sigmaA_s$ vs. frequency $\omega_s$. Red squares: KCaF$_3$. Blue dots: Silicon. The black lines denote moving averages with a window of 50 datapoints.
  	Inset: Joint probability density plot for the Gamma-point optic mode $\omega (\vec 0, 11) = 2.81$\,THz with $\sigmaA_s > 1$.}
  	\label{fig:modes}
\end{figure}

To estimate the importance of anharmonicity in a specific vibrational mode $s$, we evaluate Eq.\,\eqref{eq:sigmaAx} for the mode resolved forces $F_s$ obtained from the eigenvectors $\vec e_s$ as given by Eq.\,\eqref{eq:Fs}.
The mode-resolved degree of anharmonicity $\sigmaA_s$ is plotted in Fig.\,\ref{fig:sigma_mode_freq} as a function of the mode frequency $\omega_s$ for silicon and KCaF$_3$ at 300~K.
Across the whole vibrational spectrum, silicon exhibits almost the same mild anharmonicity of $\sigmaA_s \lesssim 0.2$.
Conversely,  KCaF$_3$, exhibits larger values of $\sigmaA_s$ with a significantly increasing magnitude and width for frequencies below 7~THz.
Below frequencies of 5\,THz, the anharmonic contributions make up for roughly 50\,\% of the forces, with several modes approaching or even exceeding $\sigmaA_s = 1$, as one of the $\Gamma$-point optical modes with $\omega_s = 2.81$\,THz. This implies that the harmonic model predicts forces that are not even qualitatively correct. They may have a significantly incorrect value and even a wrong sign. In the
joint probability density shown in the inset, the anharmonic distribution~$p(F_s^\mathrm{A})$ is thus broader than the force distribution for this particular mode.

Qualitative insight into the microscopic mechanism underlying anharmonicity can be obtained by inspecting
the displacements associated with the modes of interest. For example, the $\Gamma$-point mode with $\sigmaA_s > 1$ highlighted in the inset of Fig.\,\ref{fig:modes}
participates in the phase transition to the cubic structure above $\simeq 550$\,K~\cite{Bulou1980}.
This shows that the system begins to ``feel''  the onset of the phase transition already at 300\,K, well below the actual phase transition temperature. This aspect,~i.e.,~the relation between $\sigmaA_s$ and phase transition temperatures, is discussed further in more detail and for a broad set of systems in Secs.\,\ref{sec:DynamicallyStabilized} and \ref{sec:HT.Perovskites}.
Accordingly, the proposed measure is not only a valuable quantification and classifcation tool, but it is also sheds light on the microscopic mechanisms driving anharmonicity.

\section{Application to Dynamically Stabilized Systems}
\label{sec:DynamicallyStabilized}
\begin{figure}[h!]
	\centering
	\includegraphics{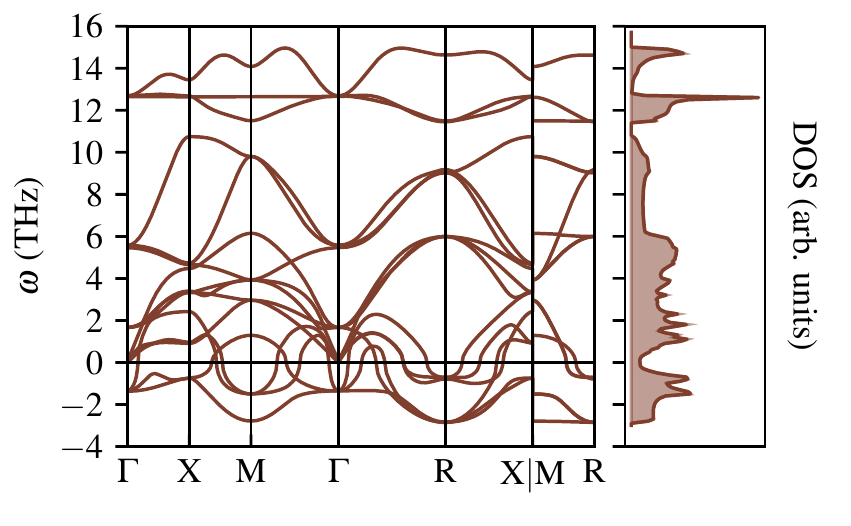}
	\caption{Phonon dispersion of cubic aristotype KCaF$_3$ with significant imaginary modes.}
	\label{fig:dispersion_KCaF3_cub}
\end{figure}

As mentioned before, KCaF$_3$
undergoes a second-order phase transition to the cubic aristotype structure above 560\,K~\cite{Bulou1980}.
This structure, which corresponds to an alignment of the octahedra (see Fig.\,\ref{fig:KCaF3}), is not a local minimum of the potential-energy surface, but a saddle point.
This can be seen from the respective phonon dispersion, which features several imaginary modes, as shown in Fig.\,\ref{fig:dispersion_KCaF3_cub}.
We use this aristotype phase of KCaF$_3$ to exemplify the meaning of $\sigmaA$ for dynamically stabilized systems,
since this is a commonly observed stabilization mechanism in strongly anharmonic materials~\cite{Errea2011, Hellman2011, Carbogno2014}, especially in perovskites~\cite{Lee2016, Saidi2016}.

\begin{figure}[h!]
	\centering
	\includegraphics{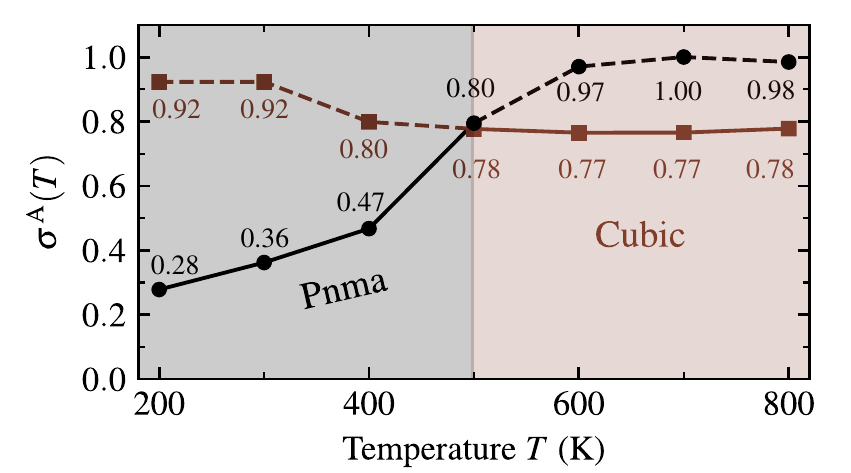}
	\caption{\label{fig:sigma_imag}
		$\sigmaA$ as a function of temperature, computed with respect to the stable orthorombic configuration (Pnma) and the dynamically stabilized cubic configuration. The shaded areas denote where the average atomic positions observed during the simulation correspond to the Pnma or cubic phase. The experimental phase transition from Pnma to cubic takes place at 550-560\,K~\cite{Bulou1980}.}
\end{figure}

Imaginary phonon frequencies as observed in Fig.~\ref{fig:dispersion_KCaF3_cub} imply a breakdown of the harmonic approximation, since
displacements from equilibrium are not energetically bounded. Accordingly,
imaginary modes cannot be used to assess a physically meaningful dynamics and
are thus typically neglected in standard perturbative methods.
With respect to the potential-energy surface, however,
a parabolic expansion around the cubic equilibrium and the evaluation of force constants is still possible.
Accordingly, the proposed definition of the $\sigmaA$-measure remains meaningful, as demonstrated in the following.
For this purpose, we
compare $\sigmaA_\t{cubic}(T)$, computed with the force constants of the cubic structure, to $\sigmaA_\t{Pnma} (T)$,
computed using the force constants of the low-temperature orthorombic Pnma phase, as shown in Fig.~\ref{fig:sigma_imag}.
At low temperatures, at which the Pnma phase is stable, $\sigmaA_\t{Pnma}$ starts from $0.28$ and increases roughly linearly up to~$\sigmaA_\t{Pnma} (400\,\t{K}) = 0.47$. Between 400 and 600\,K,
a superlinear increase of $\sigmaA_\t{Pnma}$ up to a value around $\simeq 1$ above 600\,K is observed.
Essentially, this means that forces obtained from the harmonic Pnma model are irrelevant for the nuclear dynamics at $T>600$~K. This is in line with the finding that the material undergoes a phase transition to the cubic phase at these temperatures, as also observable in the MD simulations.

For the exact same reason, the $\sigmaA_\t{cubic}$ values obtained using the harmonic model of the cubic structure are very high~$>0.9$
below the phase transition temperature, since the orthorombic Pnma structure and not the cubic one is thermodynamically stable at these conditions.
The fact that  $\sigmaA_\t{cubic}$ and $\sigmaA_\t{Pnma}$ cross each other at
$\approx$~500K~($\sigmaA=0.79$) indicates that the measure $\sigmaA$ is not only able to quantify the strong anharmonic effects active under these conditions,
but that it is also able to qualitatively capture the underlying phase transition.
As discussed in more detail in Sec.~\ref{sec:HT.Perovskites} for more perovskites, the crossing observed between $\sigmaA_\t{cubic}$ and $\sigmaA_\t{Pnma}$ is qualitatively correlated with the occurrence of the phase transition. However, the almost exact quantitative agreement between phase transition temperature and $\sigmaA$ crossing observed for KCaF$_3$ appears to be coincidental and further work is needed to understand the phenomenon in detail.

\section{Accelerating Anharmonicity Quantification}
\label{sec:Sampling}
In the previous sections, accurate but computationally involved aiMD simulations were used to explore the potential-energy surface for performing anharmonicity quantification. For scanning through material space in a \textit{high-throughput} fashion, it is desirable to obtain reliable estimates for~$\sigmaA$ at a more moderated cost,~i.e.,~by a fast computation in order to decide whether a full aiMD calculation is necessary to model the nuclear dynamics, or if
the harmonic approximation (with or without further perturbative corrections) might suffice.

For this purpose, we note that the thermodynamic averages entering
the anharmonicity metric given in Eq.\,\eqref{eq:sigmaA} and~\eqref{eq:sigmaAx}
can be evaluated approximately by sampling with the harmonic Hamiltonian,~i.\,e.,
\begin{eqnarray}
\langle O \rangle_T
&=& \frac{1}{\mathcal Z_\mathcal{V}} \int \t d \vec R \, \t e^{- \beta \mathcal V (\vec R)} ~ O (\vec R) \\
&\approx&
\langle O \rangle_T^{(2)} =
\frac{1}{\mathcal Z_{\mathcal{V}^{(2)}}}
\int \t d \vec R \, \t e^{- \beta \mathcal V^{(2)} (\vec R)} ~ O (\vec R)~,
\label{eq:thermodynamic_average_approx}
\end{eqnarray}
with $\beta = 1/k_{\rm B} T$.
For the practical evaluation of Eq.~(\ref{eq:thermodynamic_average_approx}), we generate
atomic configurations via~\cite{West2006}
\begin{align}
\Delta R_{I}^\alpha
= \frac{1}{\sqrt{M_{I}}} \sum_{s} \zeta_s \mean{A_s} e_{s I}^{\alpha}~,
\label{eq:samples1}
\end{align}
in which $\vec e_s$ are the harmonic eigenvectors, $\mean{ A_s } = {\sqrt{2 k_\t{B} T}}/{\omega_s}$ is the mean
mode amplitude in the classical limit~\cite{Dove}, and  $\zeta_s$ is a normally distributed random number~\cite{West2006}.
\begin{figure}[h!]
	\centering
	\includegraphics{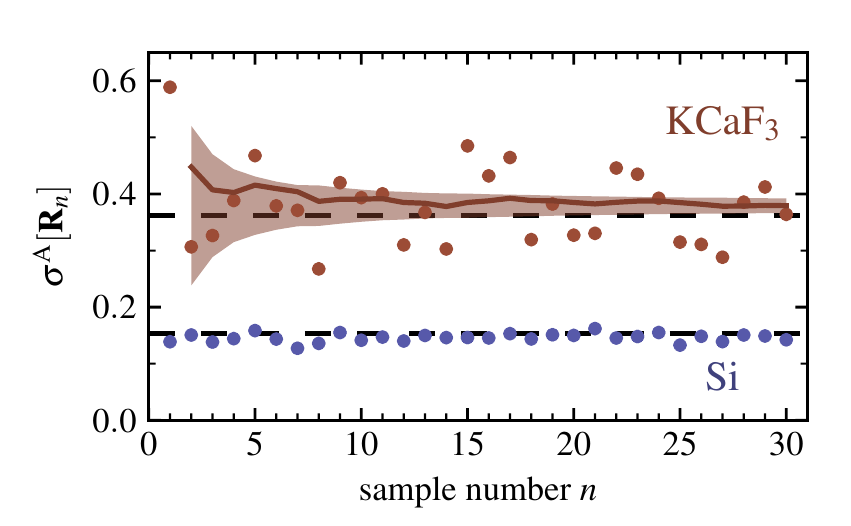}
	\caption{ \label{fig:convergence_sigma}
		Convergence of the anharmonicity measure $\sigmaA$ with respect to number samples obtained from Eq.\,\eqref{eq:samples1}. Dots: $\sigmaA [{\bf R}_n]$ for individual samples; Red line: Cumulative average.
		Black dashed line: $\sigmaA$ from \emph{ai}MD.
		Shadowed region: Convergence estimated by standard error.
	}
\end{figure}

To estimate the convergence of $\sigmaA$ with respect to the number of samples generated by Eq.\,\eqref{eq:samples1}, we evaluate $\sigmaA$ as defined in Eq.\,\eqref{eq:sigmaA} for each individual sample ${\bf R}_n$ and denote this value by $\sigmaA [{\bf R}_n]$. The values are plotted in Fig.\,\ref{fig:convergence_sigma} for a total of 30 samples.
For silicon, each of the individual samples is sufficient to obtain an estimation of $\sigmaA$ within more than 99\,\% accuracy, given that the harmonic approximation in Eq.\,\eqref{eq:samples1} holds in this case.
In the case of KCaF$_3$, where the harmonic approximation is not expected to yield a reliable dynamics, the described approach yields
a value of $\sigmaA = 0.38$ that differs from the one obtained by aiMD~($\sigmaA=0.36$) by $5\,\%$, as shown in Fig.\,\ref{fig:convergence_sigma}.
Furthermore, we find that a good estimate of $\sigmaA = 0.39 \pm 0.09$ can be obtained with less than 10 samples, thus with a computational cost
that is reduced by up to two orders of magnitude with respect to a full aiMD. Note that the estimated value of $\sigmaA$ obtained via sampling
also allows to judge how reliable the sampling itself is, as reflected by the fact aiMD and harmonic sampling coincide for Si, but differ in the case
of KCaF$_3$.

\begin{figure}[h!]
	\centering
	\includegraphics{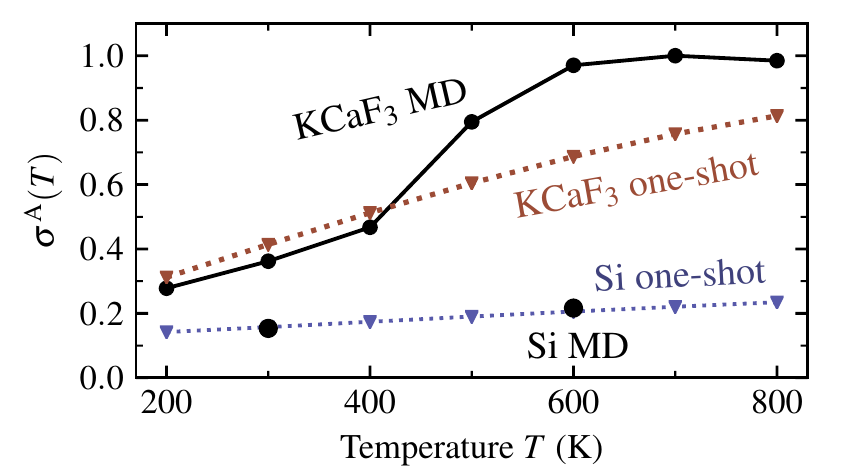}
	\caption{\label{fig:sigma_temp_oneshot}
		$\sigmaA$ as a function of temperature obtained from MD~simulations (black circles) and one-shot sampling (triangles connected by dashed curves) according to Eq.\,\eqref{eq:samples1}.
	}
\end{figure}

Exploiting this rational allows to speed up this statistical approach even further by assigning~$\zeta_s=(-1)^{s - 1}$, which generates a single, deterministic sample from the most probable
part of the random distribution~\cite{Zacharias2016}.
In the classical limit explored in this work, this essentially implies evaluating $\sigmaA(T)$ at the turning point of the oscillation
estimated by the harmonic model. This allows one to reliably single out very harmonic materials~($\sigmaA \leq 0.2$) within a single force evaluation, thus saving a further order of magnitude in computational cost. Most importantly, one can exclude that highly-anharmonic materials are misclassified
when~$\sigmaAOS \leq 0.2$, given that the anharmonicity is that small even at the turning point. We compare the temperature dependence of $\sigmaA (T)$ for this \emph{one-shot} sampling technique and aiMD in Fig.\,\ref{fig:sigma_temp_oneshot}.
For silicon, the one-shot sampling is found to be sufficient to obtain $\sigmaA$, as expected. Interestingly, also for KCaF$_3$, the one-shot sampling is able to estimate $\sigmaA$ at
low temperatures, but no longer yields quantitatively reliable estimates at elevated temperatures,
at which the actual dynamical behavior of KCaF$_3$ deviates significantly from the harmonic reference as discussed in the previous section.
Over the whole temperature range, the one-shot approach does however detect that strongly anharmonic effects are active and that aiMD simulations
are necessary to reliably treat its dynamics.

\begin{figure}
    \centering
    \includegraphics{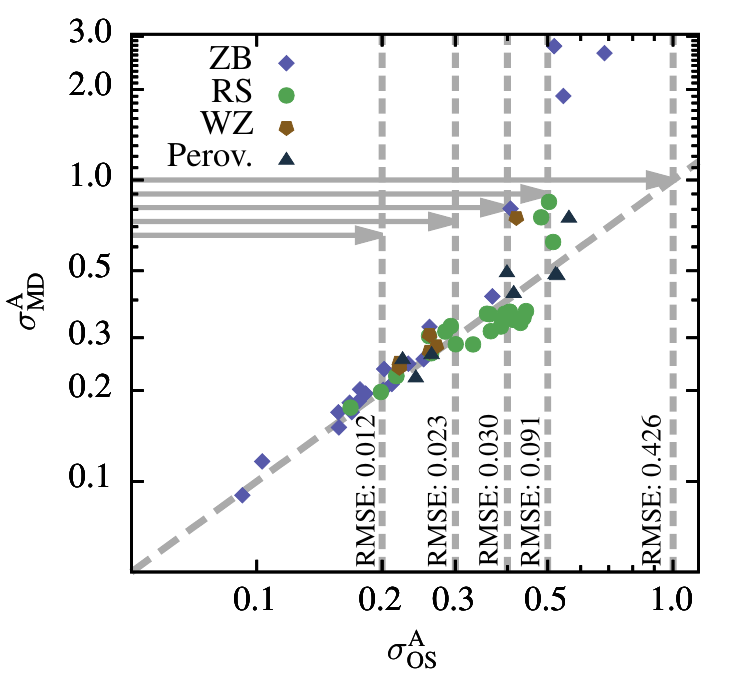}
    \caption{Comparison of $\sigmaA$ calculated from molecular dynamics and one-shot sampling for 25 rock salt, 21 zincblende, 7 wurtzite, and 10 orthorhombic perovskite materials at 300 K.}
    \label{fig:one_shot_md}
\end{figure}
This is further substantiated in Fig.~\ref{fig:one_shot_md} for a wider set of materials discussed in more detail in the following section. Here, we compare
the values of $\sigmaA$ at 300~K obtained by using the one-shot and molecular dynamics approaches with each other for 63 materials.
As expected, $\sigmaAOS$ and $\sigmaAMD$ are in good agreement with each other for Si like materials with a $\sigmaA < 0.2$, while
deviations are observed for larger values of $\sigmaAOS$, whereby the errors are more pronounced the larger $\sigmaAOS$ becomes.
In particular, we observe that $\sigmaAOS$ can yield qualitatively wrong results for $\sigmaAOS > 0.4$. For this reason,
$\sigmaAOS$ is used in the following to pre-screen the materials and to single out materials with $\sigmaA < 0.2$,
whereas aiMD is used to obtain reliable values of $\sigmaA$ whenever $\sigmaAOS > 0.2$.

\section{Application to Material Space}
\label{sec:Screening}
To substantiate and generalize the insights obtained for the two example materials in the previous section, we compute $\sigmaA$ for two distinct groups of materials at multiple temperatures: simple binary compounds (rock salts, zincblende and wurtzites) and perovskites.
For both cases, we perform symmetry-preserving geometry optimization for the structures using parametric constraints in FHI-aims until all forces are converged to a numerical precision better than \mbox{10$^{-3}$\,eV/\AA}~\cite{Lenz2019}.
From there, we calculate a converged harmonic model of each material's vibrational properties and then generate thermally displaced supercells using either molecular dynamics or the one-shot approach according to Eq.\,\eqref{eq:samples1}.
All calculations use the PBEsol functional to calculate the exchange-correlation energy and an SCF convergence criteria of $10^{-6}$ eV/\AA~and $5\times10^{-4}$ eV/\AA~for the density and forces, respectively.
Relativistic effects are included in terms of the scalar atomic ZORA approach and all other settings are taken to be the default in FHI-aims.
For all  calculations we use
the \textit{light} basis sets and numerical settings in FHI-aims.
These settings ensure a convergence in lattice constants of~$\pm 0.1~\mbox{\AA}$ and a relative accuracy in phonon frequencies of~3\%.

\begin{figure}[]
    \centering
    \includegraphics{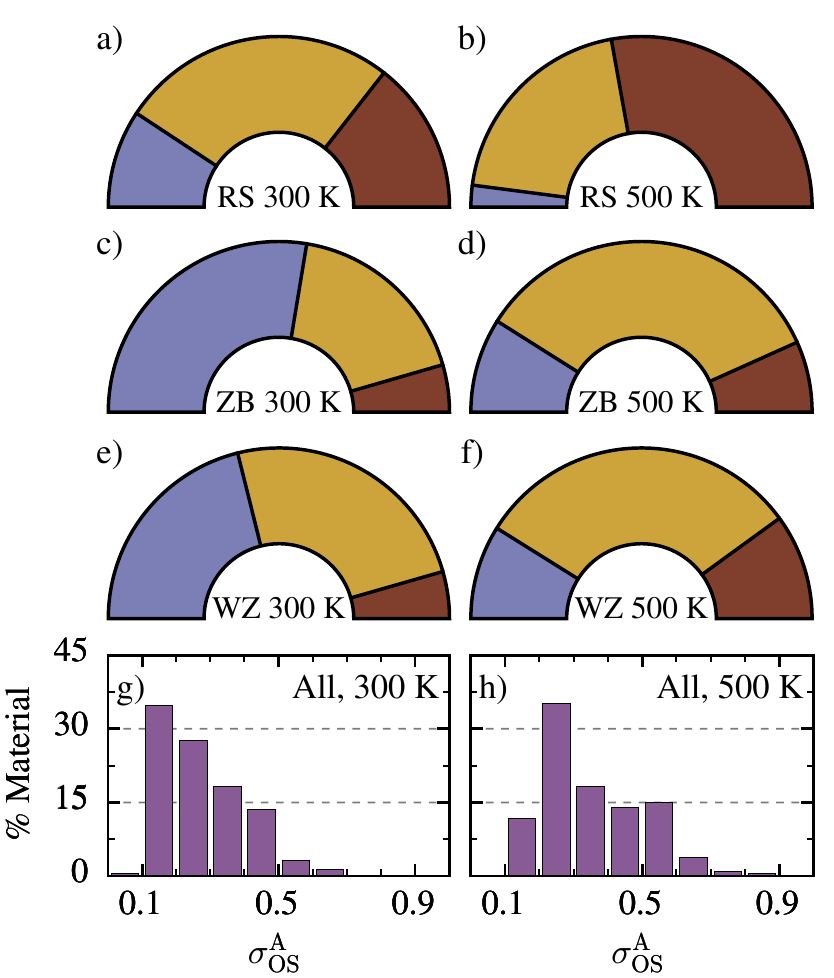}
    \caption{a-f) Pie charts representing the number of materials with a $\sigmaAOS \leq 0.2$ (blue, left), $0.2<\sigmaAOS<0.4$ (yellow, center), and $\sigmaAOS \geq 0.4$ (red, right) for a set of 97 rock salt (a, b), 67 zincblende (c, d), and 45 wurtzite (e, f) materials, at 300 K (a, c, e) and 500 K (b, d, f). The histogram for the entire set at g) 300K and h) 500 K.}
    \label{fig:sigmaA_simple_mats}
\end{figure}

\subsection{Rock salts, Zincblende, and Wurtzites}

To understand how prevalent harmonic, Si-like materials are across a broader chemical space, we use $\sigmaAOS$ to screen over
an initial test set that includes 97 rock salt (RS), 67 zincblende (ZB), and 45 wurtzite (WZ) binary and elemental solids, as summarized in Figure~\ref{fig:sigmaA_simple_mats}.
At 300 K only 35\% of all 209 materials tested can be classified as highly harmonic with a $\sigmaAOS<0.2$.
At elevated temperatures anharmonic effects get significantly stronger. At 500 K only 10\% of the materials
have a $\sigmaAOS<0.2$, while 34\%  feature a $\sigmaAOS$ value $> 0.4$. For both temperatures, zincblende
and wurtzite materials are more harmonic on average, while the majority of rock salts has a $\sigmaAOS$ value
indicative of more complicated dynamical processes.
These results are in line with experimental and theoretical studies that show materials with a higher coordination number have longer bond lengths and softer lattices leading to stronger anharmonic interactions~\cite{Miller2017a, Zeier2016}.
This screening demonstrates that anharmonic effects are more prevalent in material space than previously thought, particularly at technologically relevant temperatures.

\begin{figure}[]
    \centering
    \includegraphics{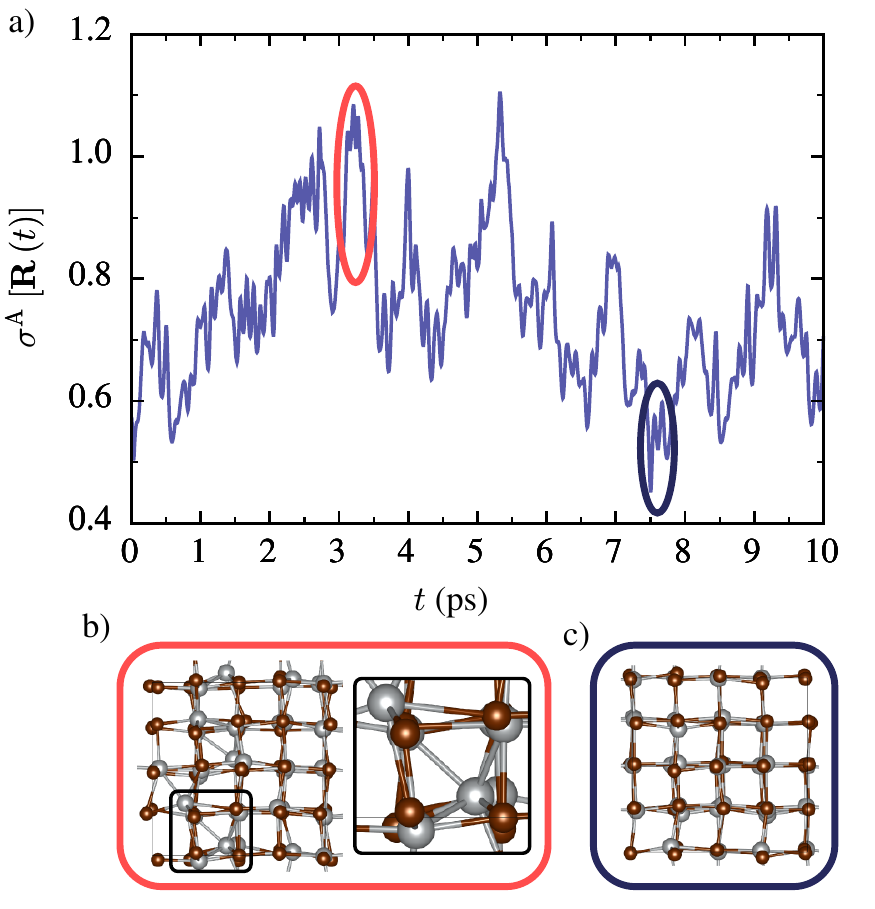}
    \caption{a) The value of $\sigmaA$ at each time step of the aiMD trajectory for rock salt AgBr. b) structure of defect (individual unitcell defects shown in the outset. c) Structure of the defect.}
    \label{fig:nmh_defect}
\end{figure}

One of the most anharmonic subclasses in the initial set of materials is noble metal halides, seven of which are among the eleven most anharmonic materials at 300 K, with only one of the remaining three materials having a $\sigmaAOS < 0.4$.
In order to analyze the suspected high anharmonicity of this class, we use aiMD~\footnote{These calculations are performed using a 64 atom supercell, the Langevin thermostat, a time step of 5 fs, and trajectory lengths of 10 ps.} to accurately calculate $\sigmaA$ for all stable noble metal halide zincblende and rock salt materials at 300~K,
as summarized in Table~\ref{tab:nmh_md_comps}. In all cases, the aiMD calculations confirm~($\sigmaAMD \ge \sigmaAOS$) the strong anharmonicity indicated qualitatively
by the one-shot approach. Quantitatively, the aiMD reveals the the actual values of $\sigmaA$ can be even substantially higher. A closer inspection of the dynamics
reveal that this is related to spontaneous defect formation, which we observe for every material except zincblende CuI. In these cases, the noble metal ions move into
the the metastable interstitial sites of the lattice forming metallic clusters within the material, as illustrated in the inset of Figure~\ref{fig:nmh_defect}b for rock salt AgBr. Similar effects have been observed in earlier aiMD studies of cuprous halides~\cite{Park1996, Bickham1999}
and have been debated extensively in literature~\cite{Gobel1996,Ulrich1999,Livescu1986}. Although a discussion of these defect-related aspects
goes beyond the scope of this work, we note that large jumps in $\sigmaA$ (i.e. $\sigmaA$ approaches or exceeds 1.0) are correlated with the
formation of defects. The jumps are a result of the harmonic model obviously failing to account for the occupation of interstitial sites, as can be seen in Figure~\ref{fig:nmh_defect}a. The actual magnitude of these $\sigmaA$-jumps is determined by the number of defects formed in the structure
and by the magnitude of the distortion from the pristine structure, whereby the occurrence of such jumps leads to strong fluctuations in
$\sigmaA$, as also summarized in Tab.~\ref{tab:nmh_md_comps}. This data reveals that these defects are particularly pronounced for
zincblende CuCl, CuBr, and AgBr, leading to $\sigmaA$ values much greater than 1.0. In this context, we would like to stress that
neither the employed trajectory length nor the used supercell size is sufficient for an accurate description of defect formation
in thermodynamic equilibrium. Nevertheless, the metric $\sigmaA$ is a useful indicator for where interesting, highly-anharmonic lattice dynamical
phenomena occur.

\begin{table}[]
	\centering
	\caption{$\sigmaA$ at 300 K values for noble metal halides in the test set as calculated from the one-shot method and aiMD. The standard deviation
        of $\sigmaA$ in aiMD trajectory is given in the last column.}
	\begin{ruledtabular}
        \begin{tabular}{ c  c  c  c c}
        	Material & \begin{tabular}[c]{@{}c@{}}Space\\ Group \end{tabular} & $\sigmaA_\mathrm{OS}$ & $\sigmaAMD$ & $\mathrm{std}\left[ \sigmaA\left[\mathbf{R}\left(t\right)\right] \right]$\\
         \hline
            AgBr   & 216  & 0.68  & 2.64 & 0.96 \\
            AgI    & 216  & 0.41  & 0.80 & 0.21 \\
            CuCl   & 216  & 0.55  & 1.90 & 0.22 \\
            CuBr   & 216  & 0.52  & 2.78 & 0.45 \\
            CuI    & 216  & 0.37  & 0.41 & 0.06 \\
            AgCl   & 225  & 0.50  & 0.85 & 0.12 \\
            AgBr   & 225  & 0.48  & 0.75 & 0.13 \\
            AgI    & 225  & 0.52  & 0.62 & 0.09 \\
        \end{tabular}
    \end{ruledtabular}
\label{tab:nmh_md_comps}
\end{table}

\subsection{Perovskites}
\label{sec:HT.Perovskites}
As discussed in Section~\ref{sec:DynamicallyStabilized} and shown in Figure~\ref{fig:sigma_imag} as KCaF$_3$ approached the cubic phase transition $\sigmaA$ quickly rose to $\sim1.0$, thereby crossing the $\sigmaA_\t{cubic}$ curve obtained with the force constants of the cubic structure.
To see how general this behavior is we calculate $\sigmaA$ at 300 K and 600 K for a set of ten perovskites using both the force constants of the orthorombic lattice and the ones obtained with atoms decorating the high-symmetry sites observed in the cubic structure.
The results for these materials, the phase transition temperatures of which span three orders of magnitude, are summarized in Table~\ref{tab:perov_md_comps}.
As the data in Table~\ref{tab:perov_md_comps} illustrates, once the perovskites are close~($\pm 150$K) to their respective transition temperatures, $\sigmaA$ tends to go above 0.45, with its overall magnitude determined by the extent of deformation away from the harmonic reference structure.
This spike also generally corresponds to a reduction of $\sigmaA_\t{cubic}$, with an apparent crossing occurring near the transition temperature\footnote{CsSnI$_3$ appears to be an exception, since $\sigmaA$ and $\sigmaA_\t{cubic}$ do not cross, but just become  comparably large after the phase transition.}.
Once past the transition temperature $\sigmaA$ does increase with increasing temperature when calculated using the cubic force constants, but generally remains below that of the orthorhomic material.
These results combined with the data from the previous sections indicate that $\sigmaA$ could be useful as an indicator for phase transitions or defect formation in materials, but further study is necessary.

\begin{table}[h!]
	\centering
	\caption{$\sigmaA$ at 300\,K and 600\,K for several perovskites along with their experimental phase transition temperatures. $\sigmaA_\text{cubic}$ is calculated from force constants of the material in the orthorhomic lattice with the atoms in the cubic high-symmetry positions.}
	\begin{ruledtabular}
        \begin{tabular}{ c  c  c  c  c  c}
        	Material & \begin{tabular}[c]{@{}c@{}}$\sigmaA$\\ (300 K)\end{tabular} & \begin{tabular}[c]{@{}c@{}}$\sigmaA_\text{cubic}$\\ (300 K)\end{tabular} & \begin{tabular}[c]{@{}c@{}}$\sigmaA$\\ (600 K)\end{tabular} & \begin{tabular}[c]{@{}c@{}}$\sigmaA_\text{cubic}$\\ (600 K)\end{tabular} & \begin{tabular}[c]{@{}c@{}}Transition\\ Temperature \\ (K)\end{tabular} \\
         \hline
            CaZrO$_3$   & 0.22 & 2.37 & 0.31 & 1.80 & 2023~\cite{Andre2014} \\
            CsCaBr$_3$  & 0.65 & 0.46 & 0.63 & 0.55 & 143~\cite{Ma2018} \\
            CsSnBr$_3$  & 0.75 & 0.65 & 0.88 & 0.77 & 247~\cite{Mori1986}\\
            CsSnI$_3$   & 0.49 & 0.82 & 0.78 & 0.84 & 351~\cite{DaSilva2015}\\
            KCdF$_3$    & 0.42 & 0.89 & 1.17 & 0.86 & 460~\cite{Yamashita1990} \\
            MgNaF$_3$   & 0.26 & 0.98 & 0.37 & 0.85 & 1038~\cite{Zhao1993}\\
            NaTaO$_3$   & 0.25 & 0.74 & 0.73 & 0.65 & 720~\cite{Kennedy1999} \\
            RbCaF$_3$   & 0.49 & 0.37 & 0.50 & 0.46 & $<$50~\cite{Flocken1986} \\
            RbCdF$_3$   & 0.48 & 0.40 & 0.54 & 0.52 & 124~\cite{Studzinski1986} \\
            SnSrO$_3$   & 0.22 & 1.02 & 0.32 & 0.81 & 905~\cite{Glerup2005}\\
        \end{tabular}
    \end{ruledtabular}
\label{tab:perov_md_comps}
\end{table}

\begin{figure}[!h]
    \centering
    \includegraphics{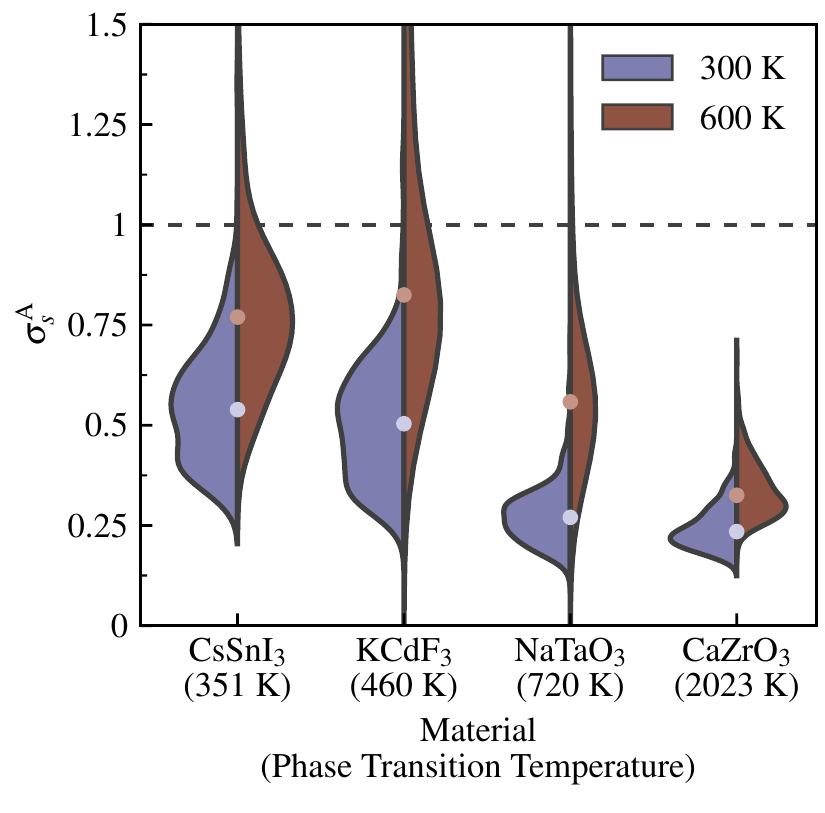}
    \caption{Violin plots of the mode resolved $\sigmaA$ for CaZrO$_3$, NaTaO$_3$, KCdF$_3$ and CsSnI$_3$ at 300 K (blue and left) and 600 K (red and right). The lighter colored circles (bottom 300\,K and top 600\, K) represent the median of the mode distribution at a given temperature.}
    \label{fig:mode_res_perov}
\end{figure}

To get a better understanding of how individual modes behave at different points near transition temperatures we illustrate the mode resolved $\sigmaA_s$ values for CaZrO$_3$, NaTaO$_3$, KCdF$_3$, and CsSnI$_3$ at 300 and 600 K using violin plots in Figure~\ref{fig:mode_res_perov}.
When the material is far from the phase-transition temperature the mode projection is similar to what was seen for Si in Figure~\ref{fig:modes} with the relative anharmonicity grouped together in a band centered around the average $\sigmaA$ value for the material, as seen for CaZrO$_3$.
As a material approaches a transition (e.g. NaTaO$_3$ at 600 K or KCdF$_3$ and CsSnI$_3$ at 300 K) the cloud broadens with a few highly anharmonic modes present.
Finally as the temperature further increases, many more modes become anharmonic as the harmonic model fails to qualitatively describe the system.
These results demonstrate the potential of this measure to not only predict when a potential phase transition will happen, but also which vibrational modes are responsible for that transition.

\section{Conclusion and Outlook}
\label{sec:conclusion}
In this work, we present a measure for the \emph{degree of anharmonicity},~$\sigmaA (T)$,
that quantifies the importance of anharmonic effects in a crystalline material.
In practice, this is done by statistically analyzing the {\it ab initio} interactions in
thermodynamic equilibrium.
This measure allows for a rapid scan through material space for the purpose of ranking materials by anharmonicity.
Our results indicate that materials whose properties are significantly affected by anharmonic effects are not uncommon at all.
Rather, largely harmonic materials like silicon and diamond with $\sigmaA < 0.2$ are the exception to the rule. In fact, only 35\% of
the binary compounds and none of the perovskites we screened over were that harmonic at room temperature. At more elevated temperatures,
anharmonic effects are even more prevalent.

The proposed metric gives access to key aspects of anharmonicity itself, since
$\sigmaA (T)$ is rigorously based on the actual interactions driving the dynamics in a material. As demonstrated for
the phase transitions occurring in perovskites, analyzing the per-mode contributions to $\sigmaA$ sheds light on the
microscopic origin of anharmonic effects that determine the macroscopic properties of materials in thermodynamic equilibrium.

\begin{figure}
	\centering
	\includegraphics{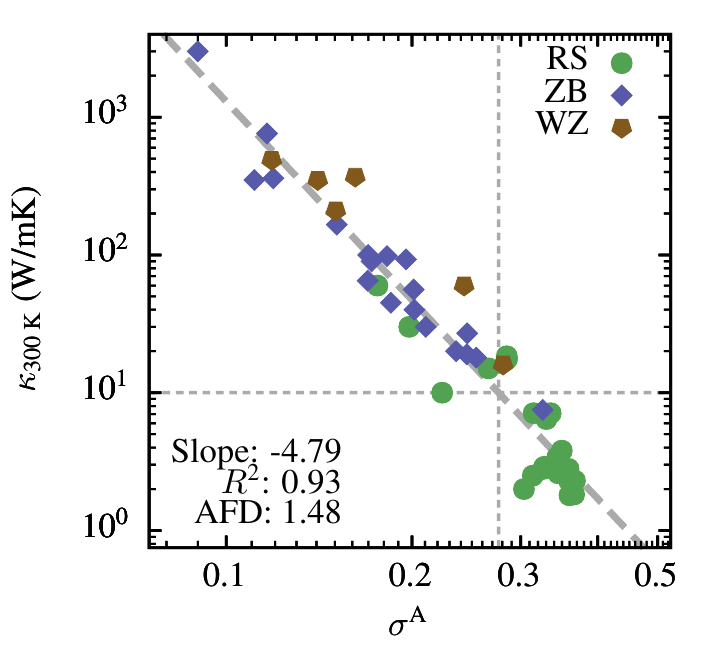}
	\caption{Comparison of $\sigmaA$ with experimental lattice thermal conductivities of 6 wurtzite, 19 zincblende, and 22 rock salt materials~(defect forming noble metal halides were excluded) with thermal conductivity values taken from Ref.~\cite{Chen2019b}.
	Values of $\sigmaA<0.2$ are obtained by one-shot sampling, values~$>0.2$ by aiMD. The diagonal gray dashed line is a power law fit of $\kappa_\text{300 K}$ with respect to $\sigmaA$. The horizontal dashed line separates materials with \mbox{$\kappa_{\rm L} < 10\,{\rm W/mK}$} and the vertical line denotes the intercept with the fit at $\sigmaA = 0.28$.}
	\label{fig:kappa_sigmaA}
\end{figure}

As an outlook, we present an additional finding in Fig.~\ref{fig:kappa_sigmaA}, in which the experimental lattice thermal conductivity~$\kappa_\mathrm{L}$ at 300 K is plotted against $\sigmaA (300\,{\rm K})$
for those RS/ZB/WZ compounds with reliable measurements of $\kappa_L$~\cite{Chen2019b}.
Fitting the data on log-log scale to a linear model, we get a slope of -4.79, implying an inverse power law between $\kappa_{\rm L}$ and $\sigmaA$, with an average factor difference (AFD) of 1.48.
AFD was introduced by Miller and coworkers to measure the accuracy of a model
via
\begin{equation}
\text{AFD} = 10^x,\, x = \frac{1}{N}\sum_{i=1}^N\left| \log{\kappa_\mathrm{L, exp}} -  \log{\kappa_\mathrm{L, model}}\right|,
\end{equation}
where $N$ is the number of samples in the test set~\cite{Miller2017a}.
The strong inverse correlation between these two properties shown in Fig.~\ref{fig:kappa_sigmaA} illustrates that $\sigmaA$ is a good descriptor for anharmonicity by itself because as a material's vibrational properties become more anharmonic, its phonon lifetimes, and therefore $\kappa_\mathrm{L}$, decrease.
It is remarkable that even without explicitly including any of the other material properties that influence $\kappa_\mathrm{L}$,
such as group velocities or heat capacities, we get a similar AFD as other semi-empirical models~\cite{Miller2017a, Chen2019b}. We note that a similar correlation is observed between $\kappa_\mathrm{L,exp}$ and $\sigmaA$ for those three perovskites in our data set, for which experimental values of $\kappa_\mathrm{L,exp}$ are available\cite{Murti1992,Qian2020,Martin1976}.
However, these few data points do not allow for a conclusive statistical assessment.
Clearly, this calls for future, more extensive research on more exhaustive data sets.

Along these lines, we like to stress again that the scope of the presented method goes beyond thermal transport and phase transition mechanisms,
and potentially applies to any phenomenon governed by anharmonic effects such as free energies~\cite{Glensk2015}, thermal expansion~\cite{Allen2020}, thermal stability~\cite{Grimvall2012}, defect formation~\cite{Glensk2014, Grabowski2011}, ferroelectricity~\cite{Poojitha2019}, and electron-phonon coupling~\cite{Zacharias2020}.
Further investigations are expected to reveal useful relationship between $\sigmaA$ and these target properties, as the thermal conductivities example above showcases.

Besides these phenomenological aspects, the findings in this work call for a systematic analysis and scrutiny of all existing, approximate treatments of vibrations in solids.
For $\sigmaA > 0.2$, the anharmonic interactions become comparable in strength to the harmonic ones, and not all phonon branches might thus be reliably described in the harmonic approximation. Furthermore, these errors propagate,~i.e.,~they affect materials properties computed on top of the harmonic approximation, for instance free energies and heat capacities as well as thermal expansion coefficients obtained via the quasi-harmonic approximation (QHA). In the latter approach, the anharmonic dependence of the force constants on the different static equilibrium positions at different volumes is accounted for, but all other anharmonic effects,~i.e., the ones arising from the actual nuclear motion, are typically neglected~\cite{Biernacki1989,Allen2015,Kim2018,Allen2020}. Indeed, a recent study shows that the QHA can severely underestimate lattice expansion in rocksalt NaBr at room temperature~\cite{Shen2019}. The fact that we find a $\sigmaA (300\,{\rm K}) = 0.40$ for NaBr suggests that a large $\sigmaA > 0.2$ can signal a breakdown of the QHA. Similarly, perturbative techniques, which are commonly used,~e.\,g., for computing lattice thermal conductivities are limited in validity for the same reasons.
In these approaches, anharmonic effects are treated by perturbation theory, starting from
the harmonic approximation and assuming the perturbation to be small~$\mathcal V^\text{A}\ll \mathcal V$.
This assumption seems to be justified for materials like silicon with $\sigmaA < 0.2$, in which anharmonic effects are responsible for 20\% of the interatomic
interactions at most. However, this assumption becomes highly questionable for the majority of materials, which, as shown in this work, exhibit $\sigmaA > 0.2$.
Especially thermal insulators generally feature large values of $\sigmaA$, cf. Fig.~\ref{fig:kappa_sigmaA}.
The formalism developed in this work is ideally suited to single out and classify well-defined test systems with different anharmonic
strength and character, ranging from simple harmonic materials~($\sigmaA \le 0.20$) up to complex materials featuring phase transitions~($\sigmaA \gg 0.2$).
Across this anharmonicity range, non-perturbative methodologies such as non-equilibrium techniques~\cite{Gibbons2009, Gibbons2011, Stackhouse2010, Puligheddu2017} and equilibrium Green-Kubo approaches~\cite{Carbogno2016, Marcolongo2016}, as well as thermodynamic integration techniques~\cite{Sugino1995,deWijs1998,Alfe2001,Glensk2015} can provide reliable benchmarks for transport and equilibrium properties, respectively, against which the various perturbative techniques at different degrees of sophistication~\cite{Broido2005, Esfarjani2011, ShengBTE, phono3py, Tadano2014, Simoncelli2019, Aseginolaza2019,Feng2017, Ravichandran2018, Puligheddu2019, Hellman2014, Tadano2015, Tadano2018, Xia2018, Eriksson2019} need to be validated.
This comparison will allow to identify up to which strength of anharmonicity~$\sigmaA$ these different techniques work
reliably and above which threshold of $\sigmaA$ perturbation theory breaks down completely.

\section{Acknowledgements}
F.K. is thankful to Olle Hellman for valuable discussions.
T.P. would like to thank the Alexander von Humboldt Foundation for their support through the Alexander von Humboldt Postdoctoral Fellowship Program.
This project was supported by TEC1p (the European Research Council (ERC) Horizon 2020 research and innovation programme, grant agreement No. 740233), BigMax (the Max Planck Society's Research Network on Big-Data-Driven Materials-Science), and the NOMAD pillar of the FAIR-DI e.V. association.
We thank the Max Planck Computing and Data Facility for computational resources.
All the electronic-structure theory calculations produced in this project are available on the NOMAD repository: \url{https://dx.doi.org/10.17172/NOMAD/2020.06.25-1}. All of the data and post-processing scripts can be found on figshare: \url{https://dx.doi.org/10.6084/m9.figshare.12783395}. Tutorials on how to generate the data with \textit{FHI-vibes}, a Python framework that works with all first and second-principles codes accessible via ASE, are available at \url{http://vibes.fhi-berlin.mpg.de/}.

\FloatBarrier
\bibliography{arxiv}

\end{document}